\documentclass{article}
\usepackage{graphicx}
\begin{document}

\title{Site testing study based on weather balloons measurements} 
\author{Aristidi E.$^1$, Agabi A.$^1$, Azouit M.$^1$, Fossat E.$^1$, Vernin J.$^1$, Sadibekova T.$^1$, Travouillon T.$^2$,\\ Lawrence J.S.$^2$,  Halter B.$^3$, Roth W.L.$^3$, Walden V.P.$^3$}
\date{$^1$ LUAN, Universit\'e de Nice, Parc Valrose, 06108 Nice Cedex 2, France\\
$^2$ School of Physics,University of New South Wales, Sydney, NSW 2052, Australia\\
$^3$ Department of Geography, University of Idaho, Moscow, Idaho, USA}
\maketitle

\section*{abstract}
We present wind and temperature profiles at Dome C measured by balloon born sonds during the polar summer. Data from 197 flights have been processed for 4 campaigns between 2000 and 2004. We show the exceptionnal wind conditions at Dome~C,  Average ground wind speed is 3.6~m/s. We noticed in mid-november the presence of high altitude strong winds (40 m/s) probably due to the polar vortex which disappear in summer. These  winds seem to have no effect on seeing measurements made with a DIMM at the same period. Temperature profiles exhibit a minimum at height 5500~m (over the snow surface) that defines the tropopause. Surface layer temperature profile has negative gradient in the first 50~m above ground in the afternoon and a strong inversion layer (5$^\circ$C over 50~m) around midnight. Wind profiles are compared with other astronomical sites, and with a meteorological model from Meteo France.
\section{Introduction}
Located at an altitude of 3250 m, on a local maximum of the Antarctic plateau, Dome C is currently the subject of an intense site testing campaign led by the University of Nice and the University of New South Wales. Already showing excellent sky transparency in the sub-millimeter (Calisse et al., 2004) and very good day and night time seeing (Aristidi et al., \cite{aris2003}, \cite{touldimm}, Lawrence et al., \cite{nature}), the site is predicted to offer better transparency and image resolution than any other known site in the world across the whole range of
 usable wavelengths. An important description of the site is contained in the wind and temperature profiles through the atmosphere. These two factors influence several aspects of telescope design and performance. Temperature and wind speed gradients result in the formation of optical turbulence. The spatial and temporal stability of these quantities in turn determine the performance of Adaptive Optics (AO) systems. Low wind speeds imply long time constants. The thickness and intensity of the turbulent ground layer usually increases with wind speed. In addition, low wind speeds imply clean optical surfaces (ice crystals are driven by surface wind) and a better overall operating environment for the mechanical systems, electronics, and observer. In this paper we present the results of wind and temperature profiles measured by balloon-borne weather sondes during four Antarctic summers between 2000 and 2004. A  statistical analysis is presented along with a comparison to other astronomical sites.

\section{Data acquisition}

The data were acquired at the Dome C station using balloon-borne weather sondes (model RS80 and RS90) manufactured by Vaisala. The sonde measures wind speed and direction by GPS triangulation as well as temperature, pressure and humidity from its in-built sensors. The humidity data will not be presented in this paper due to inconsistencies between the two types of sondes and their inaccuracies at this temperature regime. The data were taken between November and February over four seasons (2000-2004) with a good statistical coverage of all sun zenith angles. The sample used in this paper consists in a total of 197 successful balloons launches. The inversion layer, which is usually very low at Dome C, has been further sampled using captive sondes. These measurements were motivated both by the irregularity of the temperature data within the first few tens of meters of
 the balloon launches and by the importance that this part of the atmosphere has to astronomical site testing. The boundary layer is typically a major contributor to the seeing. It is therefore crucial to obtain numerous and accurate measurements of the temperature and wind profile of the low troposphere. The captive sonde temperature data were obtained by attaching the sonde to a pulley and raising it up to the top of a 30 m tower. The sonde was then slowly pulled down while the measurements were taken. As each balloon explodes at a different height, the analysis of the next sections precludes data above altitudes for which the statistical noise dominates because the number of sonde data is low. This altitude varies depending on the type of data and the time span of the particular analysis. Typically, the analysis range reaches to between 16 km and 20 km.

\section{Results}
In the following sections, the height is defined with respect to the Dome C ice level rather than the altitude above sea level (unless specified).
\subsection{Wind speed and direction}
\begin{figure}
\includegraphics[width=8cm]{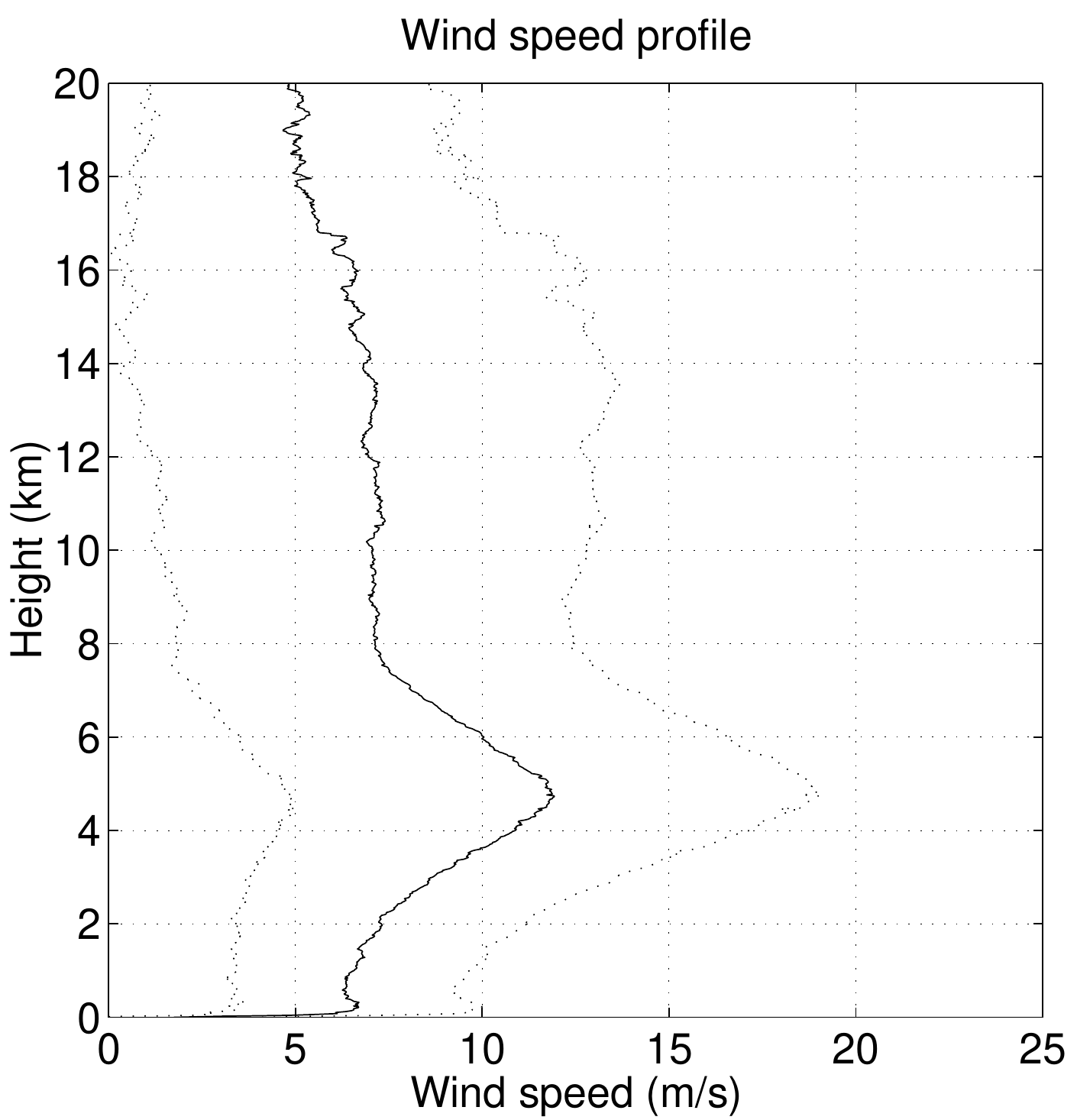}
\includegraphics[width=8cm]{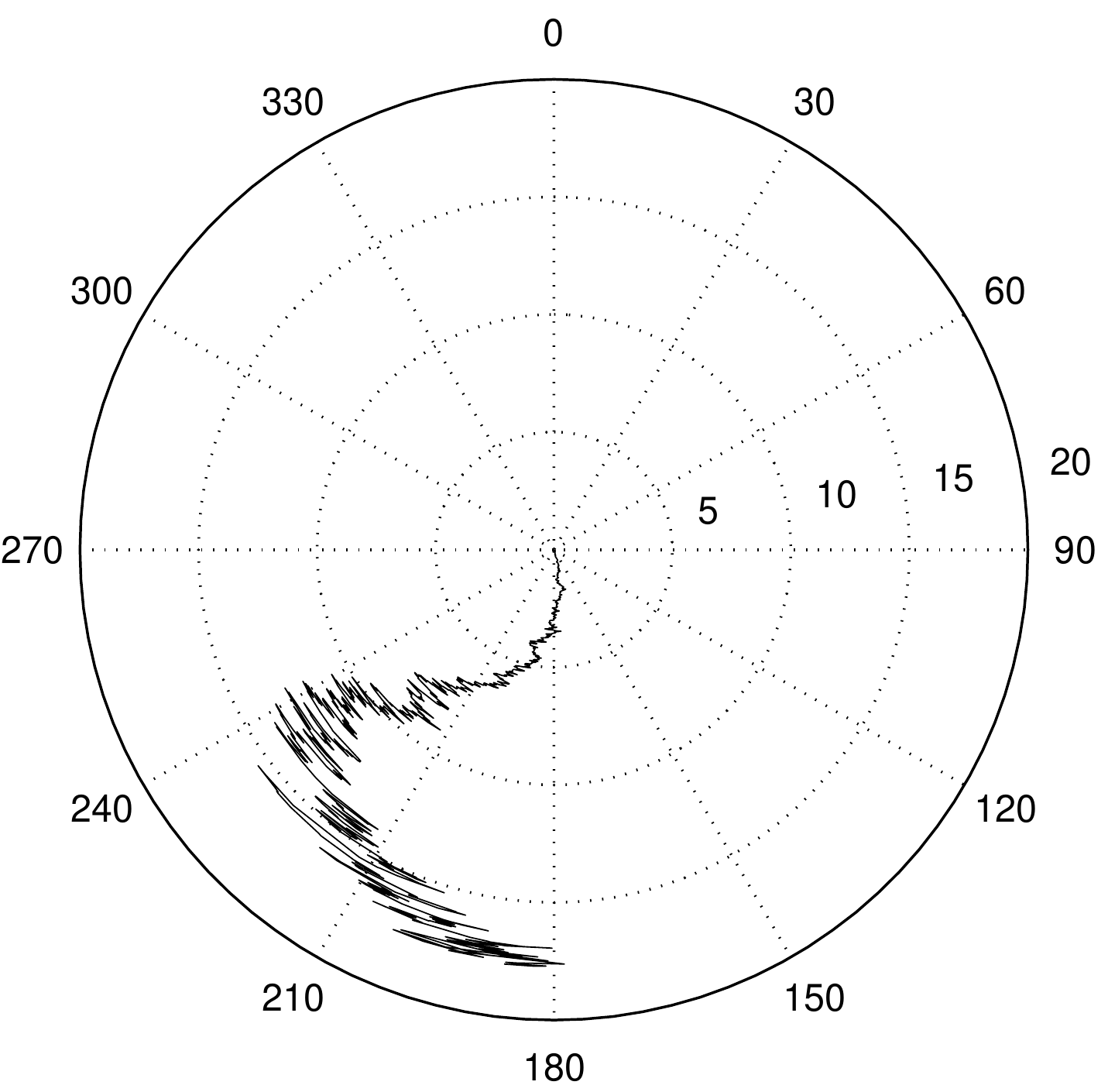}
\caption{Mean wind speed (left) and direction (right) as a function of altitude. The two outer lines on the speed profile delimit the standart deviation.}
\label{fig:winprofile}
\end{figure}

On the Antarctic continent, the wind profile is characterized by two phenomena. At the surface, katabatic winds descend from the high plateau and increase in speed as they reach the coast. Similar to inversion winds, their speed is closely related to the slope of the local terrain. At Dome C, where the slope is near zero the ground speed is very low. AWS permanent meteo station at Dome C has recorded an average value of 2.9~m/s (Aristidi et al., \cite{arisballons}).

\begin{figure}
\includegraphics[width=139mm]{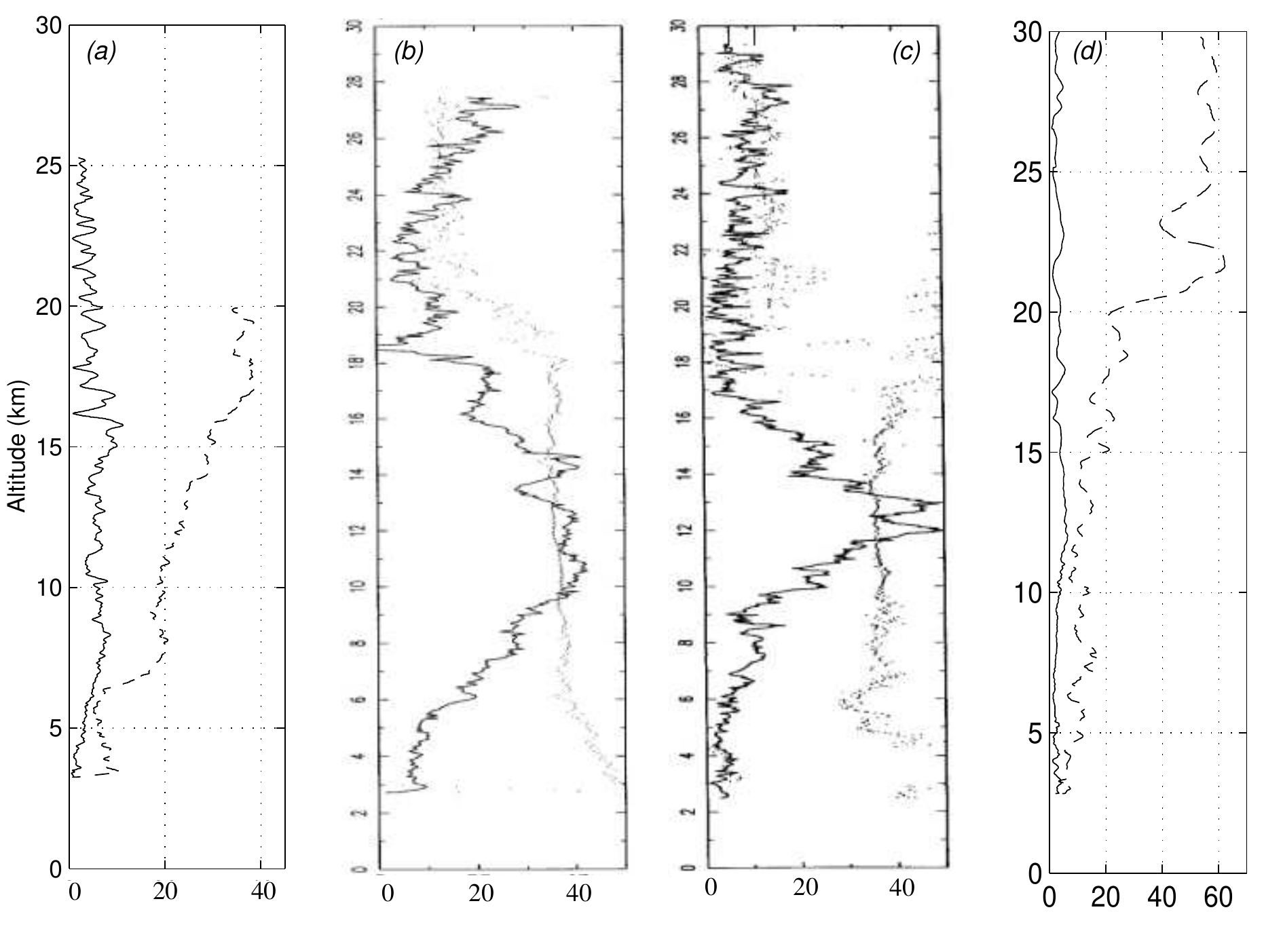}
\caption{Typical wind speed profiles at Dome C (a), Paranal (b), Mauna Kea (c) and South Pole (d). In the case of Dome C and South Pole, the dashed line is a mid-November profile while the full line is a typical summer profile. Note: the altitude is here expressed from sea level.}
\label{fig:domecparangem}
\end{figure}

 The wind speed profile (see Fig.~\ref{fig:winprofile}) is ruled by the second phenomenon that characterises the wind conditions in Antarctica, the circumpolar vortex. Arising from a large temperature gradient between the coast and the open ocean, this quasi-perfect geostrophic wind circles the Antarctic continent. In the troposphere, the seasonal variation of the vortex is minimum while in the stratosphere the vortex increases in speed in winter. This is demonstrated in Fig.~2a: while winter-time data do not yet exist, our first measurement from mid-November show speeds of up to 40 m/s at 20 km altitude. After early December the stratospheric wind drops and rarely exceeds 10~m/s.  The wind speed at this altitude is often taken as a reference altitude in site testing.  Compared to other well known astronomical sites, the wind speed profile at Dome C is very encouraging. Fig.2 shows a typical summer (full line) and end of winter (in dash) profile at Dome C and at South Pole compared with typical profiles at Mauna Kea and Paranal. In the temperate sites, the jet stream is clearly observed at an altitude of 12 km where the most intense turbulence of the free atmosphere normally occurs. The Antarctic sites have very flat summer profiles with no presence of high altitude winds. Our first measurements taken in mid-November can be interpreted as a good indication of the winter conditions. Due to the presence of the stratospheric vortex, wind speeds show a broad peak at an altitude of 20 km, substantially higher than at temperate sites.

\subsection{Temperature}
\begin{figure}
\includegraphics[width=7cm]{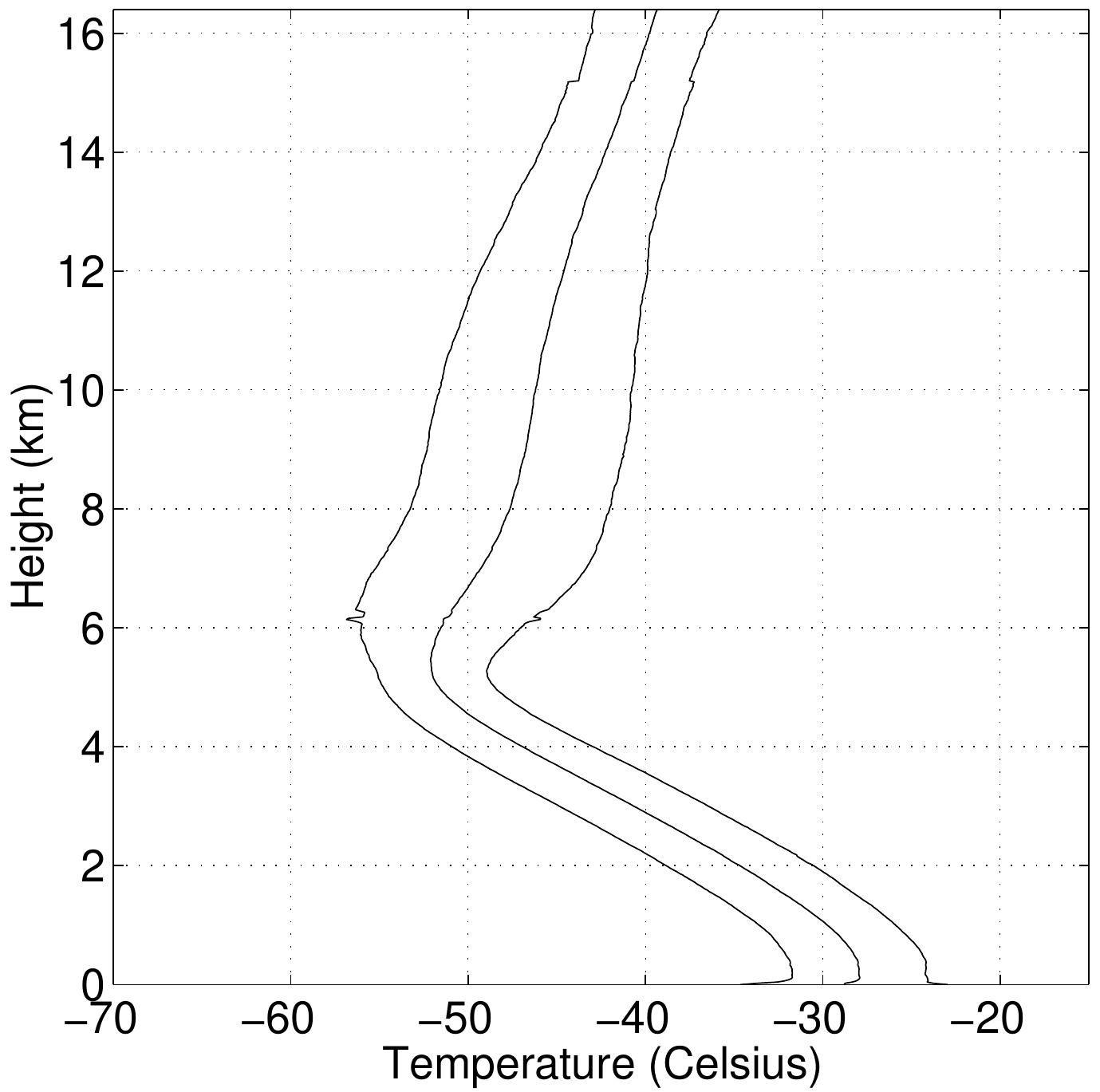}
\includegraphics[width=7cm]{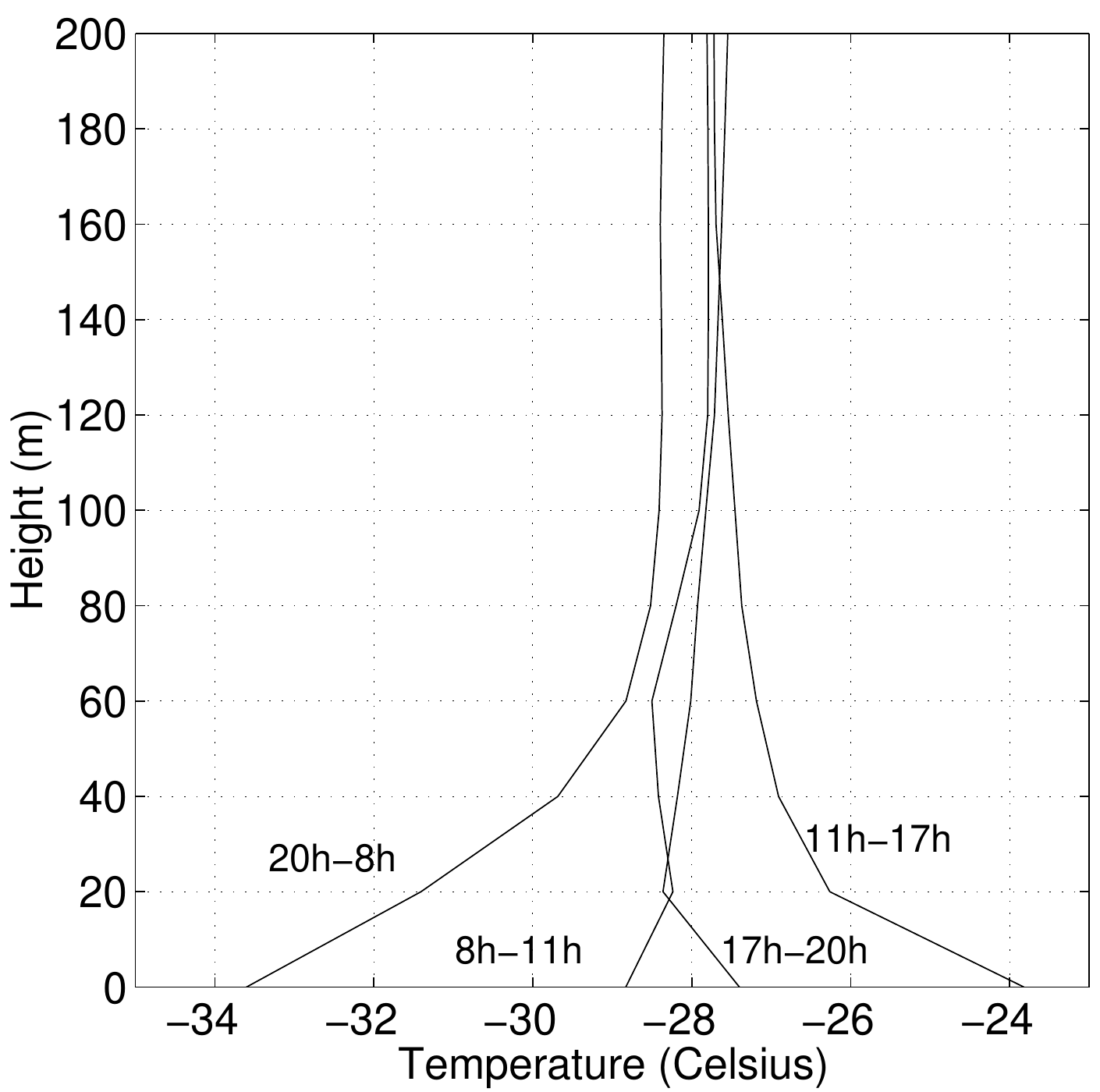}

\caption{Mean temperature profile as a function of altitude (left). The two external curves delimits the standart deviation. The second graph focuses on the first 200 m above ground for different time periods.}
\label{fig:domecparangem}
\end{figure}

The average temperature profile for Dome C is presented in Fig. 3. The tropopause, defined by the minimum temperature gradient is found at a height of 5.5 km (330 mB) above the ice and followed by a very brief isothermal layer less them 1 km wide. Fig. 3 shows also the inversion layer at Dome C in the first 200m above the snow surface as a function of time of day. In winter we expect the inversion to resemble that of the summer night time measurement with a more pronounced temperature gradient. The important point to note is the narrow depth of the inversion. Extending to only 50 m this is much lower than the boundary layer at the South Pole (220 m, Marks et al., 1999) or Vostok (300 to 500 m, King and Turner, 1997). The temperature profile in the boundary layer is almost flat twice a day, near 10~am and 5~pm; this is the period where the measured daytime seeing is the best (Aristidi~et al., \cite{touldimm}). More discussion about temperature profile and its evolution with time can be found in Aristidi~et al. (\cite{arisballons}).

\section{Comparison with meteorological models}
Figures 4 shows a comparison between meteorological models and the radio-sounding of our campaigns during the Antarctic summer 2001-2002 (wind speed, temperature and humidity, from left to right). Clearly the humidity is badly measured, badly estimated by the models or (likely) both.  The vertical sampling is of the order of 6 to 10 meters. This figure uses the atmospheric pressure for the vertical scale, so that no error can be made in altitude estimation of the radio-sounding versus the model. The general idea behind this comparison is to qualify the uncertainty of the meteorological model at Dome C, to be then able to use it during the wintertime when no radio-sounding has even been made yet.
The standard meteorological model was provided by the team of Meteo France, Toulouse. It has been obtained by a data reanalysis of the all available meteorological observations in the Southern Hemisphere at latitudes above 60$^\circ$. This model gives the atmospheric parameters for a given grid in latitude and longitude. The Dome C values were obtained by interpolation of the closest 4 points of this grid. Vertically, the model gives the parameters at 19 standard levels beginning from 1000mB to 0mB. At Dome C, the ground based altitude is 3280m, and corresponds to a pressure of about 660 mB, so that we can use only 12 to 14 points of the model for comparison, depending of the maximum altitude reached by the balloon.  The model parameters are given every six hours (0h, 6h, 12h, and 18h) for the all days of the year.  For the comparison of two data types, we use an interpolation of the models provided before and after the balloon launch time. 
Figure 4 shows that the wind speed model is quite reliable, its rms difference with the radio-sounding being less than 1 m/s. The reliability seems a little less good with the temperature, with departures of several degrees at and above the tropopause, and as mentioned, it is quite bad with the humidity parameters, both measurements and model estimations being presumably responsible. Using this fact, we can estimate the statistics of wind speed during the Antarctic winter months. This shows a consistently slow winds at any low altitudes, with a slight increase around the tropopause (300~mB), and a more important increase with speeds as fast as 40 m/s at very high altitudes, below 30 mB (Sadibekova, \cite{tanya}). These high altitude winter winds are not expected to be too damaging for the astronomical seeing because of the very low air density, while the slow winds of the lower layers are indeed a very good new for astronomers.
\begin{figure}
\includegraphics[width=15cm]{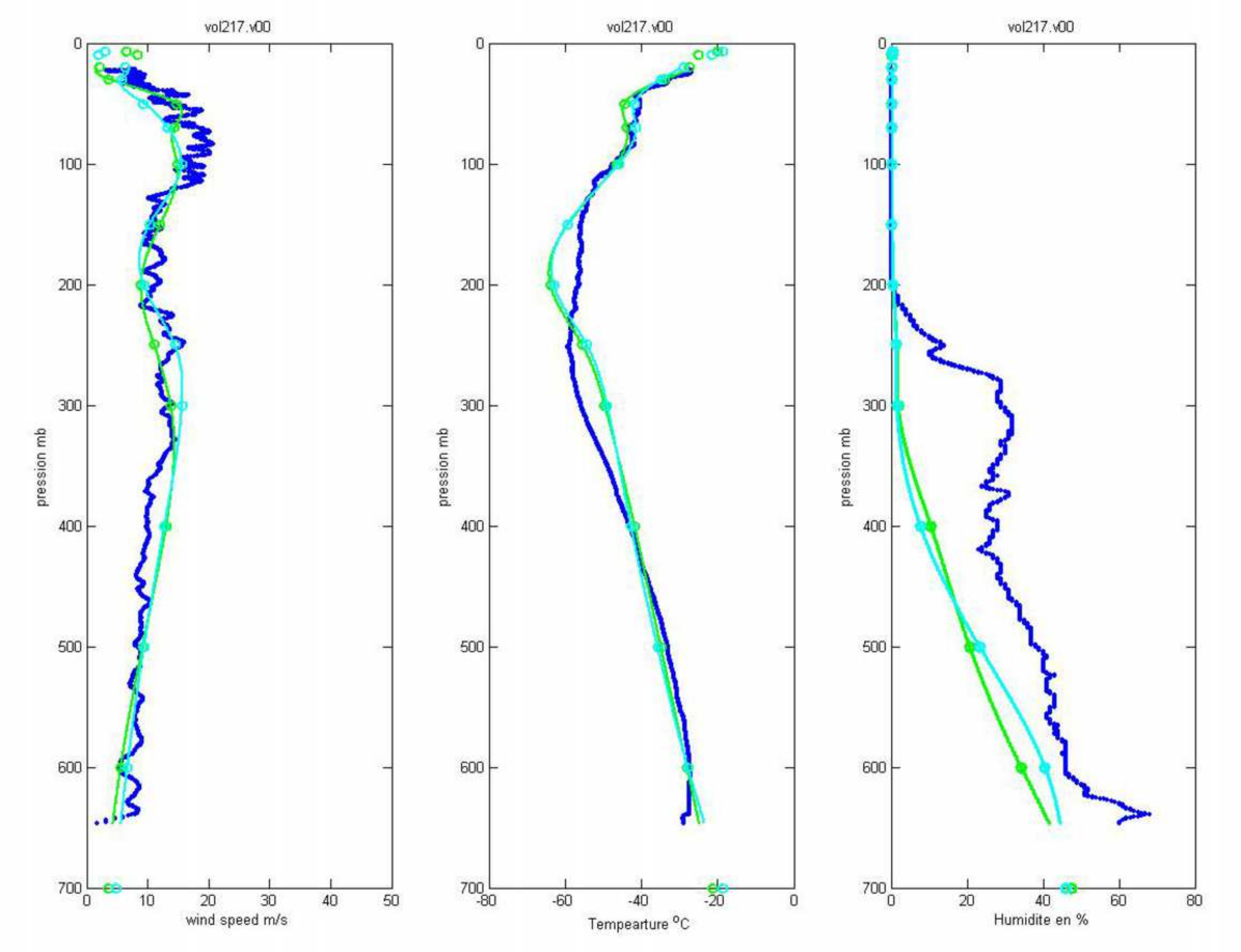}

\caption{An example (Dec. 18th, 2001) of the comparison between data model and radio-sonde measurements. The more continuous lines show the model before and after the launch, the third line is the result of the radio-sounding. }
\label{fig:domecparangem}
\end{figure}

\section{Acknowledgements}
The Concordiastro programme has been supported by a grant of IPEV. The missions on the site were supported by the logistical efforts of both IPEV and PNRA. Each one individual of these two groups deserves our most sincere thanks. The meteorogical models were provided by Paul Pettre and Fabrice Chauvin. Their contribution is really appreciated. 

\end{document}